\documentclass{article}
\usepackage{spconf,amsmath,graphicx,lipsum, hyperref}
\usepackage[utf8]{inputenc} 
\usepackage[T1]{fontenc}    
\usepackage{hyperref}       
\usepackage{url}            
\usepackage{booktabs}       
\usepackage{amsfonts}       
\usepackage{nicefrac}       
\usepackage{microtype}      
\usepackage{amsmath,amsfonts,amssymb}       
\usepackage{bm}
\usepackage{graphicx}
\usepackage{hhline}
\usepackage{xspace}
\usepackage[dvipsnames]{xcolor}
\usepackage{cite}
\graphicspath{{imgs/}}
\usepackage{epstopdf}
\usepackage{url}
\usepackage{algorithmic}
\usepackage{algorithm}
\usepackage{subfigure}
\usepackage{multirow}
\usepackage {setspace}
\usepackage{graphicx}
\usepackage{makecell}

\title{Bayesian-Boosted MetaLoc: Efficient Training and Guaranteed Generalization for Indoor Localization}

\name{Dongze Wu$^{\star}$, Jun Gao$^{\dagger}$, and Feng Yin$^{\dagger}$}

\address{$^{\star}$  Department of Statistics, University of Oxford \\ $^{\dagger}$  School of Science and Engineering, The Chinese University of Hong Kong, Shenzhen, China}

\begin{document}
\ninept 
\maketitle
\begin{abstract}
Existing localization approaches utilizing environment-specific channel state information (CSI) excel under specific environment but struggle to generalize across varied environments. This challenge becomes even more pronounced when confronted with limited training data. To address these issues, we present the Bayes-Optimal Meta-Learning for Localization (BOML-Loc) framework, inspired by the  PAC-Optimal Hyper-Posterior (PACOH) algorithm. Improving on our earlier MetaLoc~\cite{MetaLoc}, BOML-Loc employs a Bayesian approach, reducing the need for extensive training, lowering overfitting risk, and offering per-test-point uncertainty estimation. Even with very limited training tasks, BOML-Loc guarantees robust localization and impressive generalization. In both LOS and NLOS environments with site-surveyed data, BOML-Loc surpasses existing models, demonstrating enhanced localization accuracy, generalization abilities, and reduced overfitting in new and previously unseen environments.

\end{abstract}
\begin{keywords}
CSI, meta-learning, PAC-Bayesian bound, sample efficiency,
wireless localization.
\end{keywords}
\vspace*{-0.5\baselineskip}
\section{Introduction}
\vspace*{-0.5\baselineskip}
Location-based services are important in the modern society, with numerous localization methods researched across scientific domains~\cite{9992123, 6475197,yan2021graph}. While Global Navigation Satellite Systems (GNSS) excel outdoors, their performance diminishes indoors due to obstructions. As a solution, fingerprinting-based localization has gained attention for indoor settings. This involves an offline stage, where wireless signal features such as Received Signal Strength (RSS) and Channel State Information (CSI) at Reference Points (RPs) are collected to form a fingerprint database, and an online stage where Test Point (TP) signals are compared against this stored database to estimate its location using algorithms such as RADAR~\cite{bahl2000radar} and Horus~\cite{youssef2005horus}.

However, indoor wireless signal propagation faces unpredictability due to factors like multi-path effects. Even minor alteration, such as a shifting door, can induce fingerprint variations. This unpredictability complicates the creation of a consistent fingerprint database. Recent approaches leverage machine learning to tackle these issues, but they often demand extensive training data, making database creation for each setting resource-intensive and impractical for real applications. Several strategies, including data augmentation~\cite{9745151}, semi-supervised learning~\cite{chen2020semi,xing2022integrated}, and informed learning~\cite{von2021informed}, have been examined. However, most research is environment-specific~\cite{chen2017confi}, implying a singular well-performing machine learning model may not succeed universally. It is often inevitable to collect a large amount of data in the new environment, followed by updating the fingerprint database and retraining the model accordingly.

In our prior work~\cite{gao2022metaloc,MetaLoc}, we
introduced MetaLoc, which first employs meta-learning to overcome this
environment-specific limitation in localization. Specifically, the underlying localization model is
taken to be a deep neural network, and we train the meta-parameters by leveraging historical data collected from diverse well-calibrated indoor environments. These meta-parameters are then used to initialize the neural network, ensuring quick adaptation in new environments with minimal data samples. While MetaLoc significantly reduces the data requirements in the new environment, it still exhibits overfitting issues when applied to real-world datasets. Additionally, it demands tens of thousands of training tasks from the historical environments to cultivate high-quality meta-parameters, placing considerable stress on server resources. Besides, MetaLoc using a traditional neural network provides a single output for a test point, lacking the ability to quantify its prediction uncertainty. This is particularly critical in indoor localization, where complex signal propagation can lead to varying errors\cite{yang2013rssi}, emphasizing the importance of quantified uncertainty for reliable applications. To address these challenges, we enhance MetaLoc from a Bayesian perspective in this paper. Specifically, in contrast to the deterministic approach in MetaLoc, BOML-Loc assumes that the neural network parameters follow a specific distribution. By leveraging training tasks and a predefined hyper-prior, we derive an optimal hyper-posterior during the meta-training stage. This hyper-posterior serves to initialize the prior distribution of the neural network parameters in the meta-test stage. With only a few fine-tuning samples, the model can converge to the optimal posterior, resulting in a distribution of optimal network parameters for the new environment.

\vspace*{-0.5\baselineskip}
\section{Preliminaries}
\vspace*{-0.5\baselineskip}
The proposed Bayesian localization framework, BOML-Loc, leverages CSI of wireless signals for localization. CSI, a refined measurement derived from the Orthogonal Frequency Division Multiplexing wireless technology, captures multi-path characteristics of wireless channels, including signal strength and phase across subcarriers. It allows for the construction of robust fingerprints that characterize each location and enable the design of accurate localization systems~\cite{MetaLoc}.

BOML-Loc learns localization knowledge from CSI through meta-learning. Meta-learning, often termed "learning to learn," comprises two stages: meta-training and meta-test. During meta-training, the model trains a meta-learner from training tasks sampled from \(\boldsymbol{\mathcal{T}}_{train}\), aiming to capture underlying patterns across tasks. In indoor localization, CSI data from distinct domains show analogous distribution patterns but vary in noise components. We randomly select training tasks from different domains and employ meta-learning to extract shared patterns across diverse localization environments from these tasks. For \(D\) training domains (or environments), the tasks' distribution is captured as $\boldsymbol{\mathcal{T}}_{train}=\{\mathcal{T}_{1},...,\mathcal{T}_{D}\}$. 
The $i$-th training task is denoted as $\tau_{i}:= (D_{i}, S_{i})$, where $D_{i}$ denotes the domain associated with the $i$-th task out of the $D$ available domains, and $S_{i}=\{(x_{j},y_{j})\}_{j=1}^{n}$ is sampled from the distribution \(\mathcal{T}_{D_{i}}\). In the pair $(x_{i},y_{i})$, $x_{i}$ denotes the observed CSI data of dimension $52\times50\times3$, and $y_{i}$ represents the two-dimensional coordinates of the data collection position.

During the meta-test stage, the meta-learner is swiftly adapted using minimal supervised localization data from the fine-tuning dataset of a new test task, \(\tau_{test}\) sampled from \(\boldsymbol{\mathcal{T}}_{test}\), and then evaluated on its test dataset. Unlike training tasks, a test task, \(\tau_{test}\), is denoted as 
$\tau_{test}:=(D_{test},S^{0}, S_{test})$, 
where \(S^{0}\) is the fine-tuning dataset and \(S_{test}\) is the evaluation dataset. It's essential to note that the sample size in \(S^{0}\) is significantly smaller than in \(S_{test}\).

\vspace*{-0.5\baselineskip}
\section{Bayes-Optimal Meta-Learning for Localization}\label{model}
\vspace*{-0.5\baselineskip}

\begin{figure}
    \centering
    \includegraphics[scale=0.45]{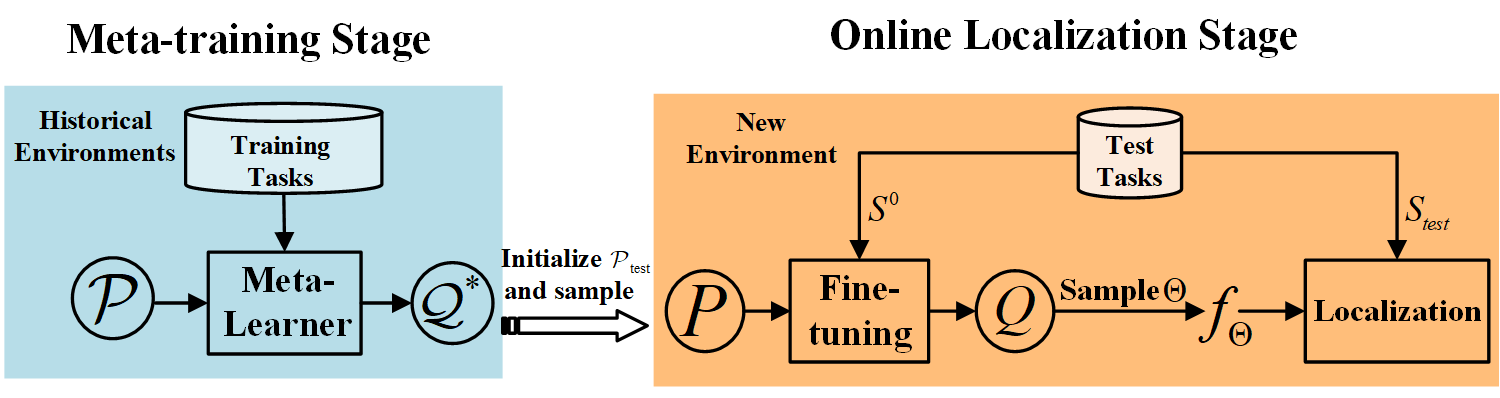}
    \caption{Overview of the proposed BOML-Loc framework with hyper-prior $\mathcal{P}$, optimal hyper-posterior $\mathcal{Q}^{*}$, prior $P$ and posterior $Q$ for neural network parameters. }
    \label{fig:overview}
\end{figure}

\subsection{Bayesian Localization Network}
\vspace*{-0.5\baselineskip}
To address the absence of test's point uncertainty estimates and the excessive data demand in most localization models, we choose a Bayesian Neural Network (BNN) as the localization network of BOML-Loc. This is due to BNN's representation of weights and parameters as distributions, which captures prediction uncertainty, and prevents overfitting by avoiding overconfident predictions\cite{goan2020bayesian}. Besides, compared to Gaussian Process or Bayesian Linear Models as utilized in recent work~\cite{PACOH} on Bayesian meta-learning, BNN excels at extracting information in high-dimensional input data\cite{cheng2022rethinking}, making it well-suited for our CSI-based localization model.

Formally, we denote the localization network by $f_{\Theta}(x)$, where $\Theta$ is the set of parameters, including the weights and biases in the localization network. We assume that $\Theta$ is drawn from a \textbf{prior distribution} $P$, specifically a Gaussian distribution, expressed as $\Theta\sim P:=N(\phi)$, where $\phi=\{\mu,\Sigma\}$, with $\mu$ and $\Sigma$ representing the mean and variance of the distribution, respectively. After observing the dataset $S$, the prior distribution $P$ is updated to a \textbf{posterior distribution} $Q$ via the Bayes rule. This process is defined as: $Q(S,P)$. 

Furthermore, to retain the robust generalization observed in MetaLoc when applied to new localization environments, we decide to employ meta-learning to train the BNN localization network. Unlike traditional BNN training which aims to learn a distribution $P$ over the model parameters $\Theta$, BOML-Loc seeks to learn a data-driven distribution over the prior $P$ by consolidating shared information from diverse localization tasks. That means, the distribution parameters $\phi$ are not fixed but are instead assumed to follow a particular \textbf{hyper-prior distribution} $\mathcal{P}$. Consequently, this introduces a corresponding  \textbf{hyper-posterior distribution} $\mathcal{Q}$ after observing the dataset.
In Fig.~\ref{fig:overview}, we present our two-stage localization framework: the meta-training stage and the online localization stage. During the meta-training stage, we aim to obtain an optimal hyper-posterior distribution \(\mathcal{Q}^*\) given localization tasks from past localization environments. This optimized \(\mathcal{Q}^*\) then serves as the initial hyper-prior \(\mathcal{P}_{0}\), from which we can sample the initial prior distribution \(P\) of the neural network parameters. By updating with a few fine-tuning data $S^{0}$ in the new environment, the prior is fine-tuned to the posterior distribution of the network parameters, denoted as $Q(S^{0},P)$. Consequently, neural network parameters $\Theta$ can be sampled from $Q(S^{0},P)$. The details are given in the following section.

\vspace*{-0.5\baselineskip}
\subsection{Meta-training Stage}
\vspace*{-0.5\baselineskip}
The goal of meta-training stage is to obtain optimal hyper-posterior $\mathcal{Q}^{*}$  using training tasks $\{\tau_{i}\}_{i=1}^{n}$ sourced from historical localization environments. To find the optimal hyper-posterior $\mathcal{Q}^{*}$, we employ the concept of the \textit{generalization-error}. In Bayes meta-learning, instead of directly finding model parameter $\Theta$ from the parameter's hypothesis space $\mathcal{H}$, the focus is now on the prior distribution $P\in \mathcal{M(H)}$, the set of all probability measures on $\mathcal{H}$. The meta-learner learns a hyper-posterior $\mathcal{Q}\in \mathcal{M(M(H))}$ from historical environments, which then serves as an initial hyper-prior $\mathcal{P}_{0}$ over prior $P$ in the meta-test stage. The hyper-posterior's localization performance in an unseen test environment is measured by the \textit{generalization-error}: 
\vspace*{-0.5\baselineskip}
\begin{small}\begin{equation}
\mathcal{L(Q,}\mathcal{\boldsymbol{T}}_{test}):=\mathbb{E}_{P\sim \mathcal{Q}}[E_{\tau_{test}\sim \mathcal{\boldsymbol{T}}_{test}}[\mathcal{L}(Q(S^{0},P),S_{test})]],
\end{equation}\end{small}
\noindent where $S^{0}$ is the fine-tuning dataset, $S_{test}$ is the test dataset, and $Q(S^{0},P)$ is the fine-tuned posterior in the new localization environment. $\mathcal{L}(Q(S^{0},P),S_{test})=\mathbb{E}_{\Theta\sim Q(S^{0},P)}\mathcal{L}(f_{\Theta}, S_{test})$ captures the loss of the localization network $f_{\Theta}$, with network parameters $\Theta$ sampled from $Q(S^{0},P)$, when applied to the test dataset $S_{test}$. 

Based on Probably Approximately Correct (PAC) learning theory\cite{PAC} and the most recent work \cite{PACOH}, we provide an upper bound on the generalization-error (1): Given a hyper-prior $\mathcal{P}$ and $\delta \in (0,1]$, we have, with probability at least $1-\delta$,
\vspace*{-0.8\baselineskip}
\begin{equation}
\begin{aligned}
\mathcal{L(Q,}\mathcal{\boldsymbol{T}}_{test}) &\leq -\frac{1}{n}\sum_{i=1}^{n}\frac{1}{\beta}\mathbb{E}_{P\sim\mathcal{Q}}[lnZ_{\beta}(S_{i},P)] \\
    &+(\frac{1}{\lambda}+\frac{1}{n\beta})D_{KL}(\mathcal{Q}||\mathcal{P})+constant,
\end{aligned}
\label{L(Q,T)}
\end{equation}

\vspace*{-0.8\baselineskip}
\noindent where $Z_{\beta}(S_{i},P)=\mathbb{E}_{\Theta\sim P}[e^{-\beta_{i}\mathcal{L}(f_{\Theta},S_{i})}]$.
The three terms of the bound represent the empirical multi-task exponential error, KL-Divergence between $\mathcal{Q}$ and $\mathcal{P}$, and a constant term, respectively. Given the training tasks $\{\tau_{1},...,\tau_{n}\}$ from various localization environments, by minimizing the upper error bound (\ref{L(Q,T)}) over $\mathcal{Q}$, along with the derivations from \cite{PACOH}, we yield the optimal hyper-posterior:
\vspace*{-0.5\baselineskip}
\begin{equation}
\begin{aligned}
  \mathcal{Q}^*(P)=\frac{\mathcal{P}(P)\exp(\frac{\lambda}{n\beta+\lambda}\sum_{i=1}^{n}\ln Z_{\beta}(S_{i},P))}{Z^{H}(S_{1},...,S_{n},\mathcal{P})}
\end{aligned}
\label{eq:Q}
\end{equation}
\vspace*{-0.5\baselineskip}

\noindent\textit{with $Z^{H}=\mathbb{E}_{P\sim\mathcal{P}}[\exp(\frac{\lambda}{n\beta+\lambda}\sum_{i=1}^{n}\ln Z_{\beta}(S_{i},P))]$.}

The PAC-Optimal Hyper-Posterior, \(\mathcal{Q}^*\), offers the best generalization guarantees over unseen localization tasks. However, considering the high-dimensionality and complexity of $\mathcal{Q}^*(P)$, direct sampling is computationally infeasible. Therefore, we introduce the Stein Variational Gradient Descent (SVGD) from \cite{SVGD}, a sampling technique that iteratively updates a set of particles to approximate high-dimensional distributions: Given a target distribution $\mathcal{Q}$, random initial particles $\boldsymbol{\phi}^{0}$, and kernal matrix $\mathbf{K}$, SVGD updates $\boldsymbol{\phi}^{0}$ via $\boldsymbol{\phi}^{j+1}\leftarrow\boldsymbol{\phi}^{j}+\eta\mathbf{K}\nabla_{\boldsymbol{\phi}^{j}}$ln$ \mathcal{Q}^*+\nabla_{\phi^{j}}\mathbf{K}$, until convergence. This results in a set $\boldsymbol{\phi}^*$ that approximates the target distribution $\mathcal{Q}$.


The update of $\boldsymbol{\phi}$ from $\mathcal{Q}^*$ via SVGD necessitates computing $\ln Z_{\beta}(S_{i},P)=\ln \mathbb{E}_{\Theta\sim P}[\exp\{-\beta_{i}\mathcal{L}(f_{\Theta},S_{i})\}]$, as shown in Eq.~\ref{eq:Q}. Its closed form can be derived for Gaussian Process-based meta-learning, as shown in \cite{PACOH}. However, our BOML-Loc, relying on a BNN, faces computational challenges due to its complicated structure. To address the challenge, we choose to employ numerical Monte Carlo estimates of \(\ln Z_{\beta}(S_{i},P)\) within the algorithm. By drawing $L$ samples  \(\{\Theta_{l}\}_{l=1}^{L}\sim P\), the estimate is:
\vspace*{-0.8\baselineskip}
\begin{equation}
    \ln \widetilde{Z}_{\beta}(S_{i},P) = \ln \left( \sum_{l=1}^{L} e^{-\beta_{i}\mathcal{L}(f_{\Theta_{l}},S_{i})} \right) - \ln L.
\end{equation}
\vspace*{-1\baselineskip}

Algorithm~\ref{al:BOML} outlines the entire meta-training process. Initially, hyper-prior particles $\{\phi_{1}^{0},\ldots,\phi_{K}^{0}\}$ are sampled from the known hyper-prior $\mathcal{P}$. At each iteration $j$, training tasks $\{\tau_{i}\}_{i=1}^{n}$ are selected from historical data. Subsequently, each particle $\phi_{k}^{j}$ is updated to approximate $\mathcal{Q}^{*}$ through SVGD, leveraging the training tasks $\{\tau_{i}\}_{i=1}^{n}$ (as seen in \textbf{Lines 5-9}). The task-level gradient 
embedding task-specific information, is evaluated in \textbf{Line 8}. This is integrated with other tasks in \textbf{Line 9}, ensuring the updating particle retains universal insights from diverse localization environments. Ultimately, the meta-training stage yields hyper-posterior particles $\boldsymbol{\phi}^*=\{\phi_{1}^*,...,\phi_{K}^*\}$ from $\mathcal{Q}^*$, which serves as initialization for the Bayesian localization network in a new environment.

\vspace*{-0.5\baselineskip}
\subsection{Online Localization Stage}
\vspace*{-0.5\baselineskip}

Due to the unpredictable nature of wireless signal propagation in a novel localization environment, preliminary fine-tuning of $\boldsymbol{\phi}^*$ is essential to capture the distinctive patterns of the new setting before actual testing. In the fine-tuning step, we initialize $\boldsymbol{\phi}^{0}$ as the $\boldsymbol{\phi}^*$ trained from the meta-training stage, and thus initialize a set of well-trained initial prior distributions $\{P_{\phi_{1}^{0}},...,P_{\phi_{K}^{0}}\}$ over the model parameters $\Theta$. Given a test task $\tau_{test}=\{D_{test},S^{0},S_{test}\}$, $\boldsymbol{\phi}^{0}$ is fine-tuned with $S^{0}=\{(x_{i},y_{i})\}_{i=1}^{r}$ from the new environment, following \textbf{Line 14-17}, to transform $\{P_{\phi_{1}^{0}},...,P_{\phi_{K}^{0}}\}$ into a set of fine-tuned posteriors $\{Q(S^{0},P_{\phi_{1}^{0}}),...,Q(S^{0},P_{\phi_{K}^{0}})\}$, represented as $\{Q_{\phi_{1}^*},...,Q_{\phi_{K}^*}\}$. This set of fine-tuned posteriors provides performance guarantees on unseen localization tasks, as it optimizes the generalization-error bound (2).

For real localization activities in a new environment, deterministic localization networks must be sampled. To optimize the localization knowledge encapsulated within each posterior particle's distribution, we choose to sample multiple localization networks per particle and aggregate their results. In the localization testing step, for each particle $\phi_{i}^*$, we sample a set of model parameters $\{\Theta_{j}^{(i)}\}_{j=1}^{N}$ from $Q_{\phi_{i}^*}$ (\textbf{Line 20}). Each point $(x_{m},y_{m})$ is tested against $\{f_{\Theta_{j}^{(i)}}\}_{j=1}^{N}$ (\textbf{Line 21}), resulting in an average error $\epsilon_{m}^{(i)}=\frac{1}{N}\sum_{j=1}^{N}||f_{\Theta_{j}^{(i)}}(x_{m})-y_{m}||_{2}$. We then compute the localization error $\epsilon_{m}$ of a single test point $(x_{m},y_{m})$, as the average of $\epsilon_{m}^{(i)}$ across all particles. This yields $\epsilon_{m}=\frac{1}{K}\sum_{i=1}^{K}\epsilon_{m}^{(i)}$. Finally, the localization error for the test dataset $S_{test}=\{(x_{m},y_{m})\}_{m=1}^{n_{test}}\sim \boldsymbol{\mathcal{T}}_{test}$ is calculated as $\frac{1}{n_{test}}\sum_{m=1}^{n_{test}}\epsilon_{m}$.


\begin{algorithm}[ht]
\caption{Bayes-Optimal Meta-Learning for Localization}
\begin{algorithmic}
\STATE \textbf{Underlying Localization Model:}
\STATE A regression Bayesian Neural Network
\STATE \textbf{Require:} Sampled hyper-prior particles $\boldsymbol{\phi}^{0}=\{\phi_{1}^{0},...,\phi_{K}^{0}\}$ from hyper-prior $\mathcal{P}$; SVGD Kernel function $k(\cdot,\cdot)$; step size $\eta$ 

\STATE \textbf{Meta-training Stage (in the historical environments):}
\STATE 1. Gather CSI data \( x \) at each reference point and log its coordinates \( y \) across \( D \) dynamic training environments
\STATE 2. Construct \( \boldsymbol{\mathcal{T}}_{train}=\{\mathcal{T}_{1},...,\mathcal{T}_{D}\} \) from $D$ environments
\STATE =======Start to train the Bayesian localization network======
\STATE 3. While not converged do:
\STATE 4. \hspace{0.3cm} Sample training tasks $\{\tau_{i}\}_{i=1}^{n}\sim\boldsymbol{\mathcal{T}}_{train}$
\STATE 5. \hspace{0.3cm} for $\phi_{k}^{j}$ in $\{\phi_{1}^{j},...,\phi_{K}^{j}\}$ do:
\STATE 6. \hspace{0.6cm} for $\tau_{i}$ in $\{\tau_{1},...,\tau_{n}\}$ do:

\STATE 7. \hspace{0.9cm} ln$\widetilde{Z}_{\beta}(S_{i},P_{\phi_{k}^{j}})\leftarrow \ln \sum_{l=1}^{L} e^{-\beta_{i}\mathcal{L}(f_{\Theta_{l}},S_{i})}-$ln$L$
\STATE 8. \hspace{0.6cm} $\nabla_{\phi_{k}^{j}}$ln$\mathcal{Q}^*\leftarrow \nabla_{\phi_{k}^{j}}$ln$\mathcal{P}+\frac{\lambda}{\lambda+n\beta}\sum_{i=1}^{n}\nabla_{\phi_{k}^{j}}$ln$\widetilde{Z}_{\beta}(S_{i},P_{\phi_{k}^{j}})$
\STATE 9. \hspace{0.3cm} $\boldsymbol{\phi}^{j+1}\leftarrow\boldsymbol{\phi}^{j}+\eta\mathbf{K}\nabla_{\boldsymbol{\phi}^{j}}$ln$ \mathcal{Q}^*+\nabla_{\boldsymbol{\phi^{j}}}\mathbf{K}$
\STATE 10. return $\boldsymbol{\phi}^*$ when it converges
\STATE
\STATE \textbf{Online Localization Stage (in the new environment):}
\STATE =======Fine-tuning the Bayesian localization network=======
\STATE 11.  Initialize $\boldsymbol{\phi}^{0}$ as the $\boldsymbol{\phi}^*$ from Meta-training Stage
\STATE 12. Sample a test task $\tau_{test}\sim\boldsymbol{\mathcal{T}}_{test}$
\STATE 13. While not converged do:
\STATE 14. \hspace{0.3cm} for $\phi_{k}^{j}$ in $\{\phi_{1}^{j},...,\phi_{K}^{j}\}$ do:
\STATE 15. \hspace{0.6cm} ln$\widetilde{Z}_{\beta}(S^{0},P_{\phi_{k}^{j}})\leftarrow  \ln \sum_{l=1}^{L} e^{-\beta_{i}\mathcal{L}(f_{\Theta_{l}},S^{0})}-$ln$L$
\STATE 16. \hspace{0.6cm} $\nabla_{\phi_{k}^{j}}$ln$\mathcal{Q}^*\leftarrow \nabla_{\phi_{k}^{j}}$ln$\mathcal{P}+\frac{\lambda}{\lambda+\beta}\nabla_{\phi_{k}^{j}}$ln$\widetilde{Z}_{\beta}(S^{0},P_{\phi_{k}^{j}})$
\STATE 17. \hspace{0.3cm} $\boldsymbol{\phi}^{j+1}\leftarrow\boldsymbol{\phi}^{j}+\eta\mathbf{K}\nabla_{\boldsymbol{\phi}^{j}}$ln$ \mathcal{Q}^*+\nabla_{\boldsymbol{\phi^{j}}}\mathbf{K}$
\STATE 18. return $\boldsymbol{\phi}^*$ when it converges
\STATE ===============Localization Testing================
\STATE 19. For $\phi_{i}^*$ in $\{\phi_{1}^*,...,\phi_{K}^*\}$ do:
\STATE 20. \hspace{0.3cm} Sample localization networks $\{f_{\Theta_{j}^{(i)}}\}_{j=1}^{N}\sim Q_{\phi_{i}^*}$
\STATE 21. \hspace{0.3cm} Test $\{f_{\Theta_{j}^{(i)}}\}_{j=1}^{N}$ against $S_{test}=\{(x_{m},y_{m})\}_{m=1}^{n_{test}}$
\STATE 22. Report localization error
\end{algorithmic}
\label{al:BOML}
\end{algorithm}

\vspace*{-0.5\baselineskip}
\section{Experimental Results}\label{experiment}
\vspace*{-1\baselineskip}

\begin{table}[htbp]
\caption{Localization results in the new environment}
\centering
\scalebox{0.66}{
\begin{tabular}{ccccccccc} 
\toprule[1.5pt] 
\multicolumn{1}{c}{\multirow{2}*{\textbf{Methods}}}& \multicolumn{3}{c}{\textbf{Line-of-Sight}}&\multicolumn{3}{c}{\textbf{Non-Line-of-Sight}}\\
\multicolumn{1}{c}{}&\textbf{\makecell[c]{Mean-\\errors / std(m)}}&\textbf{\makecell[c]{Un-\\certainty (m)}}&\textbf{Data}&\textbf{\makecell[c]{Mean-\\errors / std(m)}}&\textbf{\makecell[c]{Un-\\certainty (m)}}&\textbf{Data}\\  
\hline 
\multicolumn{1}{c}{\textbf{BOML (ours)}}                     & \multicolumn{1}{c}{\textbf{2.01 / 1.13}}          & \multicolumn{1}{c}{\textbf{1.07}}  & \multicolumn{1}{c}{\textbf{30}}          &  \multicolumn{1}{c}{\textbf{2.91 / 1.25}}         & \multicolumn{1}{c}{\textbf{1.18}} & \multicolumn{1}{c}{\textbf{30}}       \\

\multicolumn{1}{c}{MAML}                     & \multicolumn{1}{c}{2.45 / 1.23}          & \multicolumn{1}{c}{$/$}  & \multicolumn{1}{c}{30}          & \multicolumn{1}{c}{3.23 / 1.44}         & \multicolumn{1}{c}{$/$} & \multicolumn{1}{c}{\textbf{30}}        \\

\multicolumn{1}{c}{MAML-DG}                     & \multicolumn{1}{c}{2.30 / 1.24}          & \multicolumn{1}{c}{$/$}  & \multicolumn{1}{c}{30}         & \multicolumn{1}{c}{3.18 / 1.37}        & \multicolumn{1}{c}{$/$}& \multicolumn{1}{c}{30}         \\

\multicolumn{1}{c}{TL}                     & \multicolumn{1}{c}{2.42 / 1.26}          & \multicolumn{1}{c}{$/$}  & \multicolumn{1}{c}{30}          & \multicolumn{1}{c}{3.98 / 1.97}         & \multicolumn{1}{c}{$/$}&\multicolumn{1}{c}{30}          \\

\multicolumn{1}{c}{ConFi}                     & \multicolumn{1}{c}{2.89 / 0.48}          & \multicolumn{1}{c}{$/$}  &\multicolumn{1}{c}{900}          & \multicolumn{1}{c}{3.53 / 0.47}         & \multicolumn{1}{c}{$/$}&\multicolumn{1}{c}{900}         \\

\multicolumn{1}{c}{ILCL}                     & \multicolumn{1}{c}{3.61 / 2.06}          & \multicolumn{1}{c}{$/$}  &\multicolumn{1}{c}{900}          & \multicolumn{1}{c}{3.48 / 1.62}         & \multicolumn{1}{c}{$/$}&\multicolumn{1}{c}{900}         \\

\multicolumn{1}{c}{KNN}                     & \multicolumn{1}{c}{2.73 / 1.35}          & \multicolumn{1}{c}{$/$}  &\multicolumn{1}{c}{900}     &\multicolumn{1}{c}{3.35 / 1.42}         & \multicolumn{1}{c}{$/$}&\multicolumn{1}{c}{900}         \\

\bottomrule[1.5pt]
\end{tabular}}
\label{tab-results}
\end{table}

\begin{figure}
\centering
\subfigure[All LOS Environments]{\includegraphics[scale=0.27]{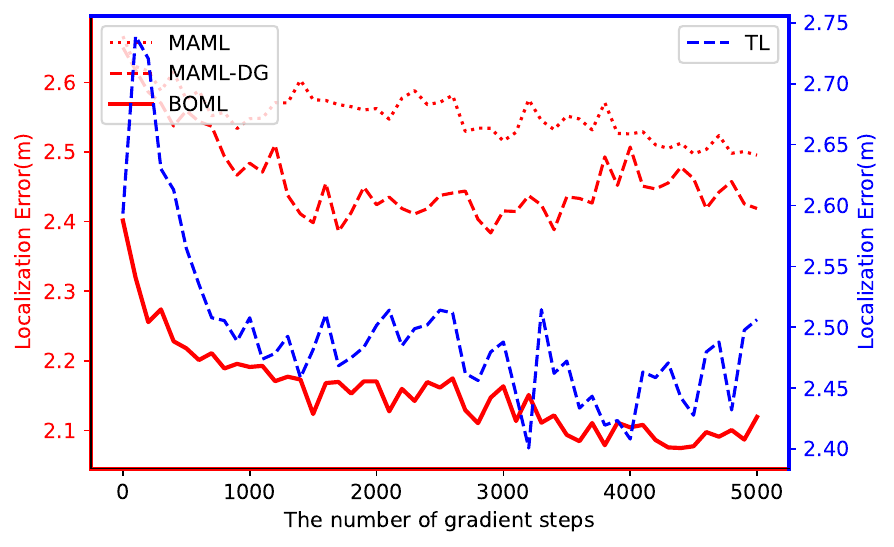}}
\subfigure[All NLOS Environments]{\includegraphics[scale=0.27]{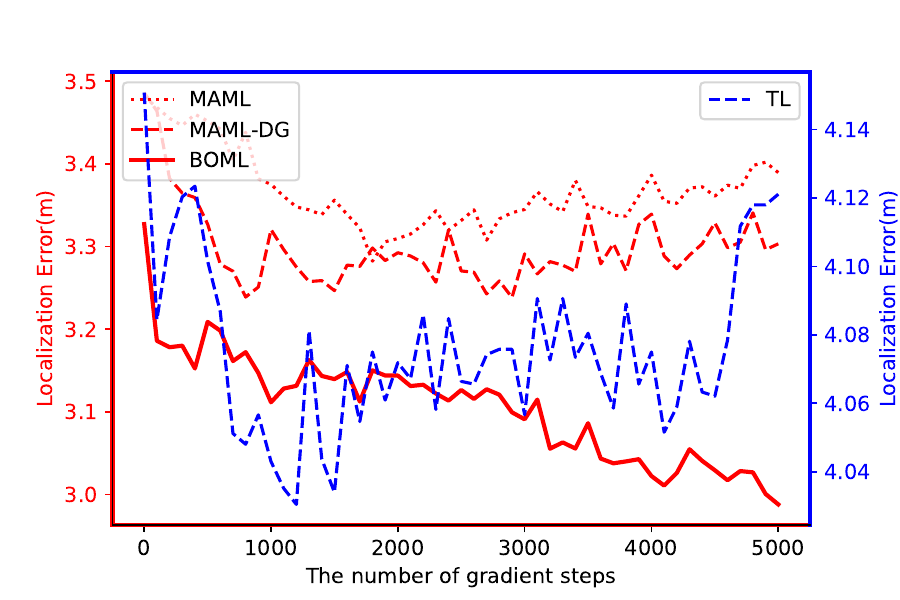}}
\subfigure[Complicated LOS]{\includegraphics[scale=0.27]{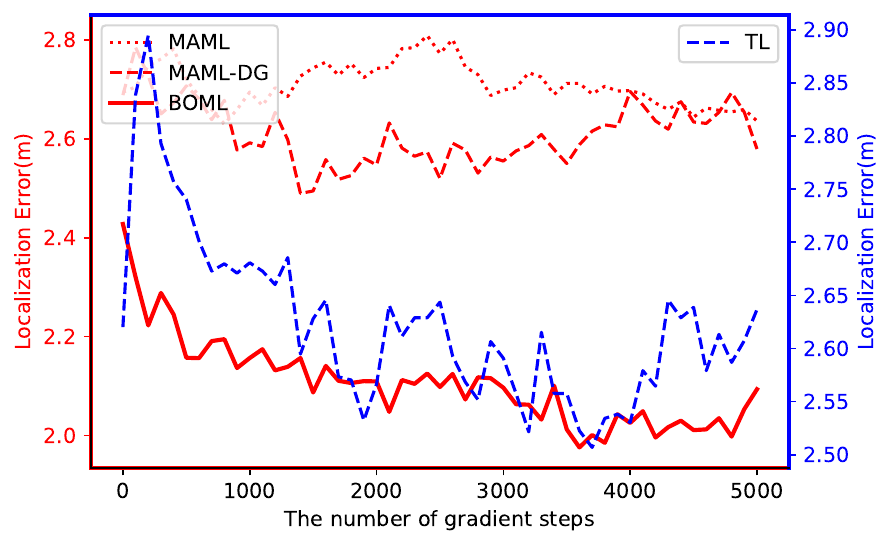}}
\subfigure[Complicated NLOS]{\includegraphics[scale=0.27]{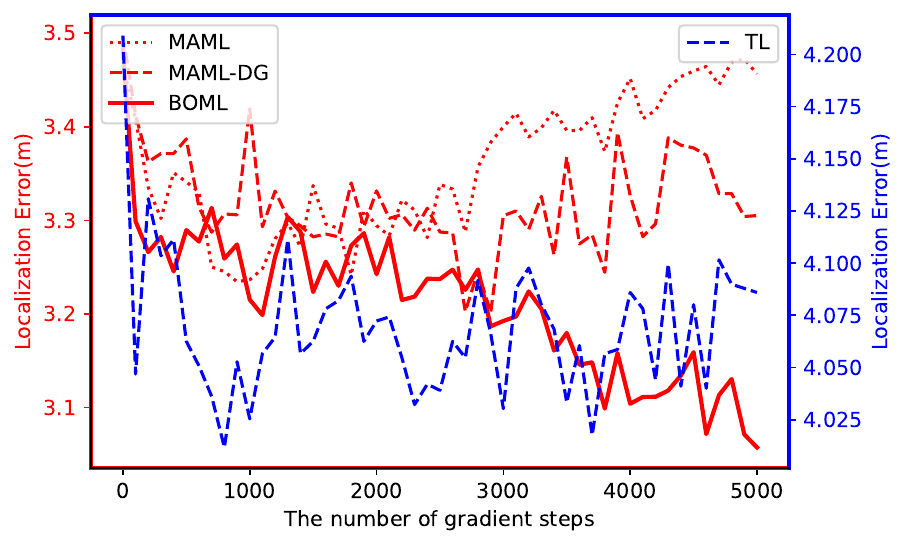}}
\caption{Convergence of averaged localization errors across datasets in both all and only complicated LOS/NLOS environments with 100 training tasks. \textbf{In
the double-axis system, the red-axis represents BOML, MAML, and MAML-DG, while the blue-axis represents TL}.}\label{fig:test-error}
\end{figure}

\begin{figure}[ht]
\centering
\subfigure[All LOS Environments]{\includegraphics[scale=0.27]{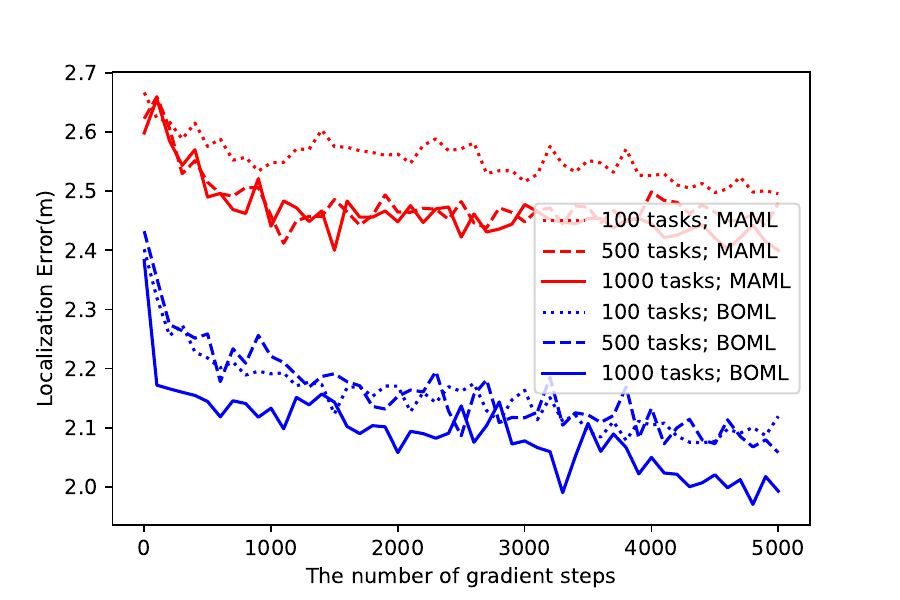}}
\subfigure[All NLOS Environments]{\includegraphics[scale=0.27]{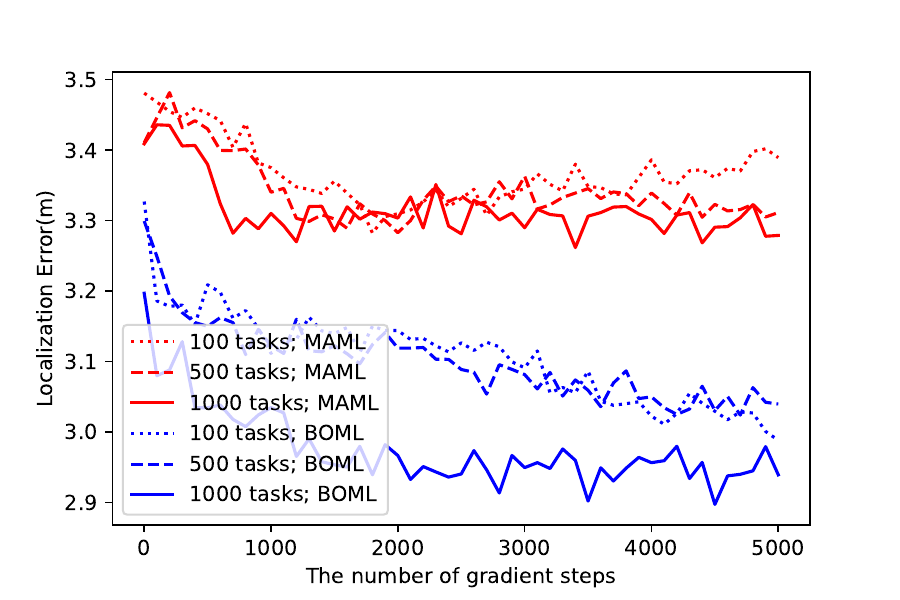}}
\caption{Convergence of averaged localization errors across datasets in all LOS/NLOS environments with different number of training tasks.}\label{fig:test-error-dates}
\end{figure}

In this section, we frame localization as a regression problem and demonstrate the effectiveness of the proposed BOML-Loc using real site-surveyed data from \cite{MetaLoc} in both Line-of-Sight (LOS) and Non-Line-of-Sight (NLOS) environments.

\vspace*{-1\baselineskip}
\subsection{Experimental Setting}
\vspace*{-0.5\baselineskip}
For each training task $\tau_{i}:=(D_{i},S_{i})$, $S_{i}\sim \mathcal{T}_{D_{i}}$ contains 50 samples. Test tasks $\tau_{test}:=(D_{test},S^{0},S_{test})$ comprise a fine-tuning dataset $S^{0}$ of 30 samples and a test dataset $S_{test}$ of 50 samples. Our base model is a BNN with four hidden layers and $Relu$ activations. For computational efficiency, diagonal Gaussian priors $P_{\phi_{k}}=\mathcal{N}(\mu_{k},diag(\sigma_{k}^2))$ are used, where $\phi_{k}:=(\mu_{k}, \ln\sigma_{k})$. A zero-centered, spherical Gaussian hyper-prior $\mathcal{P}:=\mathcal{N}(0,\sigma_{\mathcal{P}}^2)$ is applied to the prior parameter $\phi$, with $\sigma_{\mathcal{P}}$ set to 0.5. The system uses observed CSI fingerprints as inputs and predicts two-dimensional coordinates. During meta-training, parameters including learning rate, number of hyper-posterior particles, and number of posterior particles are set to 0.002, 5, and 10 respectively. Data collection spanned five days for both LOS and NLOS settings. Four days contribute to training tasks (historical environments), while a separate day is reserved for test tasks (new environments) in each evaluation. Every day among the five is subjected to testing.

\vspace*{-1\baselineskip}
\subsection{Baseline Models}
\vspace*{-0.5\baselineskip}
In this section, we assess the performance of BOML-Loc in two dimensions. Firstly, we compare BOML-Loc with state-of-the-art localization models, including MAML and MAML-DG introduced in  MetaLoc framework \cite{MetaLoc}, as well as transfer learning (TL)~\cite{MetaLoc}, KNN, ILCL~\cite{zhu2022intelligent}, and ConFi~\cite{chen2017confi}. Our primary focus in this comparison is on localization errors and generalization capabilities. Furthermore, we provide a detailed comparison between BOML-Loc, MAML, and MAML-DG, focusing on efficient training and overfitting. Through this evaluation, we aim to highlight the improvements BOML-Loc offers over potential risks observed in MetaLoc framework.
\vspace*{-1\baselineskip}
\subsection{Results}
\vspace*{-0.5\baselineskip}

Table \ref{tab-results} demonstrates the localization results of different models tested in a new environment, where BOML-Loc, MetaLoc (MAML and MAML-DG), and TL are trained using 100 training tasks, in contrast to ConFi and ILCL, which utilize 10,000 training tasks to achieve best performance. The table only displays the best localization results for each model. For models that may be susceptible to overfitting due to limited data, their localization errors are recorded prior to the start of overfitting. Notably, due to BOML-Loc's Bayesian approach, multiple deterministic neural networks are sampled from the posterior, allowing the reporting of estimation uncertainty (in terms of standard deviations in meters) for each test point. This inclusion of uncertainty provides reliability for practical localization applications.

Fig.~\ref{fig:test-error} illustrates the convergence of averaged localization errors across datasets in both all and only complicated LOS/NLOS environments, \textbf{with 100 training tasks}. Fig. \ref{fig:test-error-dates} shows the convergence for LOS/NLOS environments with training tasks set at 100, 1,000, and 10,000. Through further analysis on the localization results, we observe that BOML-Loc possesses the following advantages:
\vspace*{-1\baselineskip}
\subsubsection{Localization Performance and Generalization Abilities:}
\vspace*{-0.5\baselineskip}
From Table \ref{tab-results}, BOML-Loc, with 100 training tasks, records the lowest mean localization errors: 2.01 m in the LOS room and 2.96 m in the NLOS room. The diminished performance in the NLOS room is due to its complex environment with multiple facilities, where obstacles can impede data propagation. Notably, baseline models including MetaLoc (MAML and MAML-DG), ILCL, ConFi, and KNN all exhibit higher errors than BOML-Loc. Furthermore, BOML-Loc's performance improves with more training tasks and still surpasses MetaLoc, as illustrated in \ref{fig:test-error-dates}. 

Moreover, in new test environments, BOML-Loc requires only 30 data samples for fine-tuning to attain superior results. In contrast, ConFi, ILCL, and KNN need 900 samples for optimal fine-tuning. This demonstrates BOML-Loc's efficient generalization in dynamic and previously unseen environments, with the need of minimal human efforts in collecting fine-tuning data.



The efficient generalization of BOML-Loc can be attributed to two reasons. Firstly, using meta-learning, BOML-Loc captures shared information across fluctuating localization environments by leveraging multiple training tasks from varied historical environments, thus facilitating generalization to a new test environment. Secondly, the algorithm determines an optimal $\mathcal{Q}^*$ for the generalization error bound (\ref{L(Q,T)}), offering performance guarantees even when faced with previously unseen tasks in these ever-evolving indoor environments.
\vspace*{-1\baselineskip}
\subsubsection{Efficient Training and Over-Fitting Mitigation:}
\vspace*{-0.5\baselineskip}
As shown in Fig. \ref{fig:test-error}-Fig. \ref{fig:test-error-dates}, compared to BOML-Loc in both LOS and NLOS settings, MetaLoc (MAML and MAML-DG), exhibits higher localization errors and std, especially when the number of training tasks is limited. As reported in \cite{MetaLoc}, with 10,000 training tasks, MAML and MAML-DG can achieve the best localization results (2.11 m and 2.07 m in LOS; 3.10 m and 3.04 m in NLOS). Notably, BOML-Loc, with enough iteration steps but only 100 tasks, is even able to achieve comparable localization errors.

Notably, MAML, MAML-DG, and TL display severe overfitting issues with limited training tasks, particularly in complicated LOS/NLOS environments. In comparison, BOML-Loc consistently presents a steady performance improvement as gradient steps increase. This is because the hyper-prior $\mathcal{P}$ of PACOH acts as meta-level regularizer by penalizing complex priors that are unlikely to convey useful localization information. This property will reduce the necessity for extensive data collection across varied training environments.

\vspace*{-0.6\baselineskip}
\section{Conclusion}\label{conclusion}
\vspace*{-0.5\baselineskip}
In this study, we introduced BOML-Loc, a Bayesian meta-learning approach designed for localization in dynamic and previously unseen environments, especially with very limited training data. BOML-Loc stands out primarily for its efficient training and overfitting mitigation: Despite only trained with 100 tasks from historical environments, it outperforms state-of-the-art models including MetaLoc that use 10,000 tasks. Moreover, it showcases robust generalization capabilities, adapting efficiently to new environments with few fine-tuning data. Additionally, its Bayesian framework enables per-test-point uncertainty estimation, ensuring reliability for real-world localization applications. Given its gradient-based foundation, BOML-Loc is versatile and applicable to diverse classification and regression models, offering insights for broader signal processing challenges.

\bibliographystyle{IEEEbib}
\bibliography{strings,refs}

\end{document}